\documentclass[epj]{svjour}

\usepackage{epsfig,graphicx}

\newcommand{\Gfour}{{\sc Geant4 }}

\begin{document}

\title{ Simulation of diffraction dissociation in quark-diquark representation } 
\author{ 
V.M.~Grichine\thanks{ 
visitor at CERN, e-mail:  Vladimir.Grichine@cern.ch}
}
\institute{
Lebedev Physical Institute, Moscow, Russia 
}

\date{Received: date / Revised version: date}

\abstract{
A low mass single diffraction dissociation mode based on the quark-diquark 
representation of hadron-nucleon interaction is proposed. The mass spectra of nucleon-pion system are 
compared with experimental data and predictions of other models. The  hadron-nucleus 
single-diffraction cross-sections are calculated in the framework of the Glauber-Gribov 
model for integral cross-sections. The model predictions are compared with experimental 
data for the  different distributions of secondary particles produced in the hadron-nucleus 
interactions in the momentum range 31-320 GeV/c. 
}

\maketitle

\section{Introduction}
\label{intro}

\begin{figure}
\includegraphics[height=2.8in,width=3.5in]{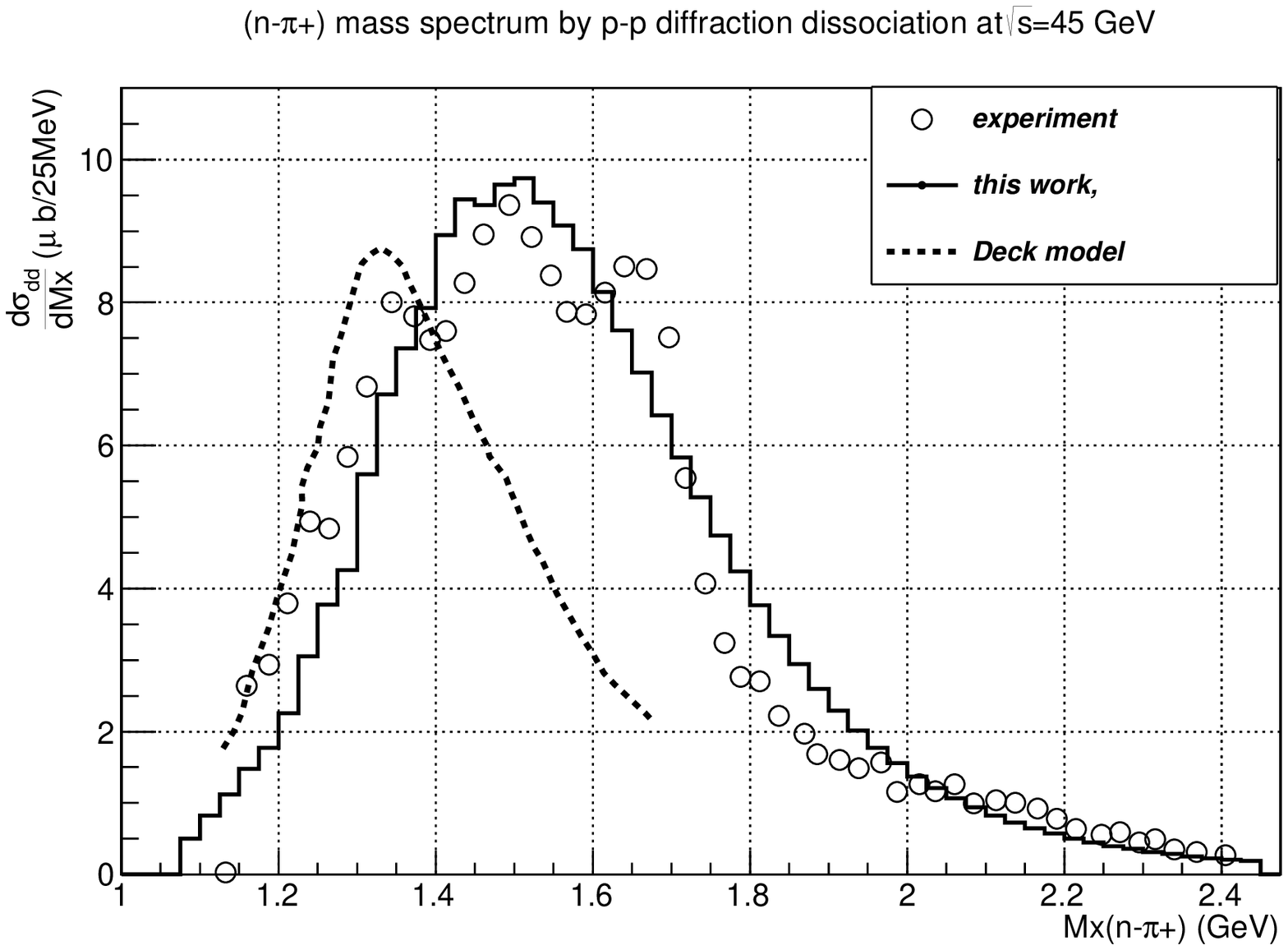}
\caption{ The invariant mass spectrum of the proton-pion system produced in the proton-proton interactions 
at $\sqrt{s}=45$~GeV. The solid histogram corresponds to calculations according to this work, 
the dashed line is the Deck-model calculations from~\cite{gerhold1976}, the points are 
experimental data~\cite{kerret1976}.}
\label{mass2}
\end{figure}

Diffraction dissociation (DD) or inelastic diffraction scattering has more than 60 years history. 
It was proposed by E.L.~Feinberg and I.Ya.~Pomeranchuk in 1953 and became widely known due to 
their seminal review paper in 1956~\cite{fp1956}.   
A short time later, Good and Walker~\cite{gw1960} proposed the general approach to DD based on the quantum 
field framework. These ideas were transformed to a working model by Drell and Hiida~\cite{dh1961} 
and Deck~\cite{d1964}. The resulting Drell-Hiida-Deck (DHD) model was broadly used for the 
description of experiential data, see review~\cite{zt1988}. The DHD-model was capable to 
describe the main features of DD, however the low mass spectra of excited hadrons are not 
correctly predicted by the DHD model (see Fig.\ref{mass2} and the references in there).

In this paper, a low mass single diffraction dissociation model based on the quark-diquark 
representation of hadron-nucleon interaction is proposed. The mass spectra of the nucleon-pion system are 
compared with experimental data and the predictions of other models. The  hadron-nucleus 
single-diffraction cross-sections are calculated in the framework of the Glauber-Gribov 
model for integral cross-sections. The model predictions are compared with experimental 
data for the  different distributions of secondary particles produced in the hadron-nucleus 
interactions in the momentum range 31-320 GeV/c.


\begin{figure}
\includegraphics[height=2.8in,width=3.5in]{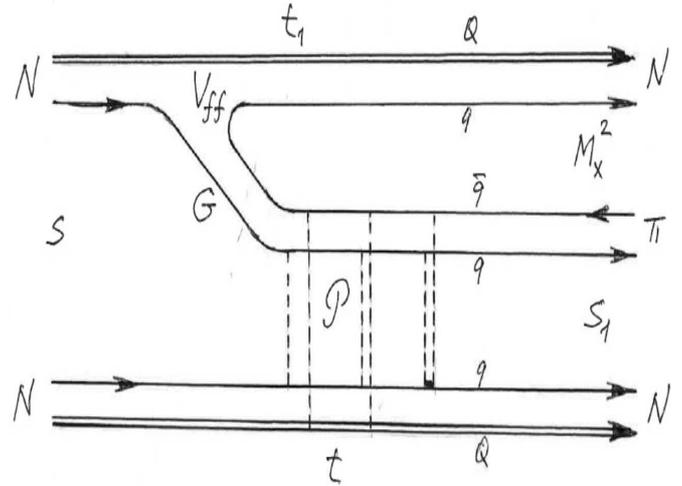}
\caption{ The quark-diquark diagram for the diffraction dissociation of nucleon $N$ in 
the nucleon-nucleon interaction. $M_x^2$ is the invariant mass of the excited nucleon squared, 
$t$ is the momentum transfer between nucleons.}
\label{diagram}
\end{figure}

\begin{figure}
\centering \includegraphics[width=3.5in,height=2.8in]{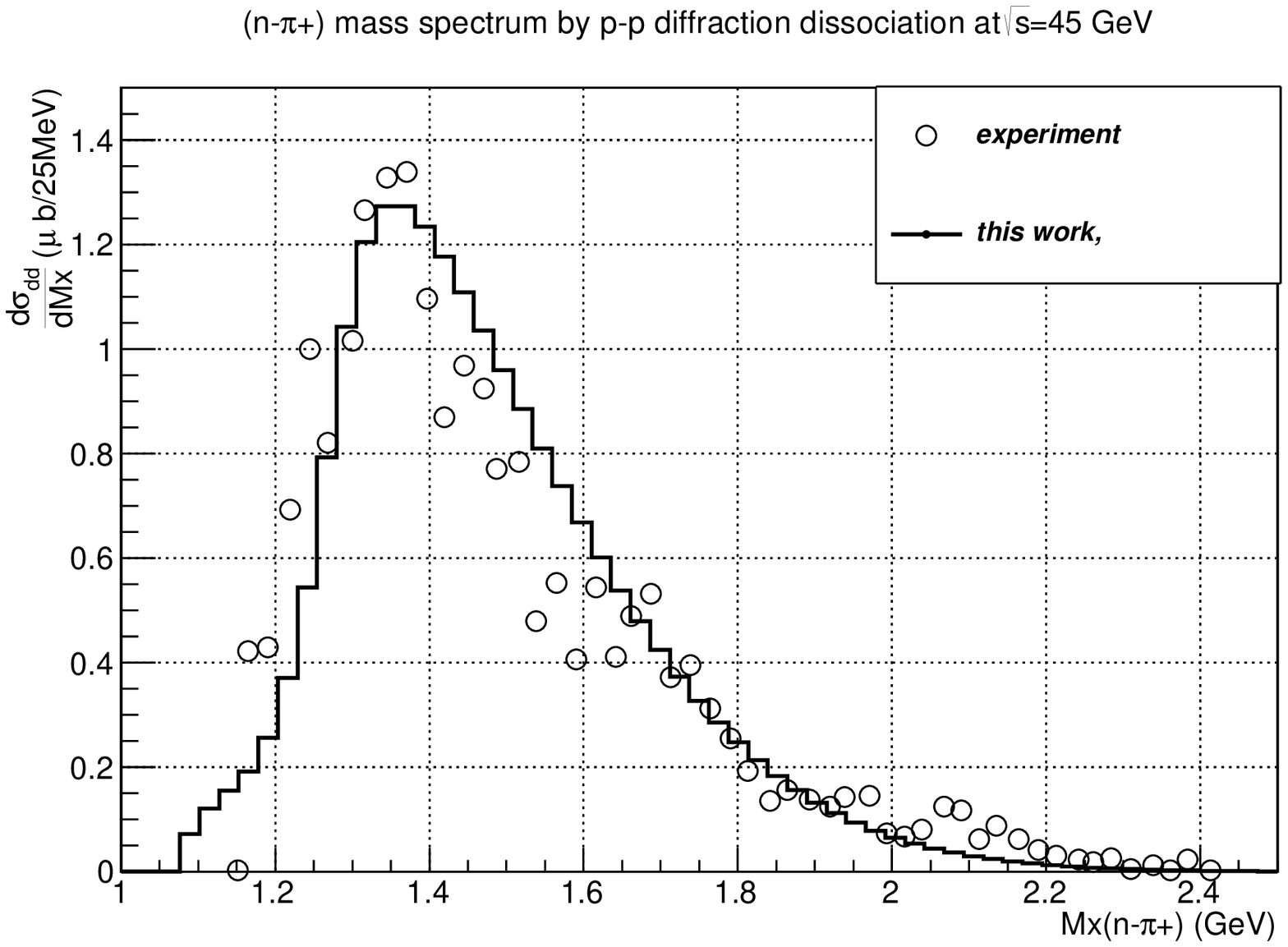}
\caption{ The invariant mass spectrum of the proton-pion system produced in the proton-proton interactions 
at $\sqrt{s}=45$~GeV. The histogram corresponds to calculations according to this work, the  
points are experimental data~\cite{kerret1976}.}
\label{mass1}
\end{figure}

\begin{figure}
\includegraphics[height=2.8in,width=3.5in]{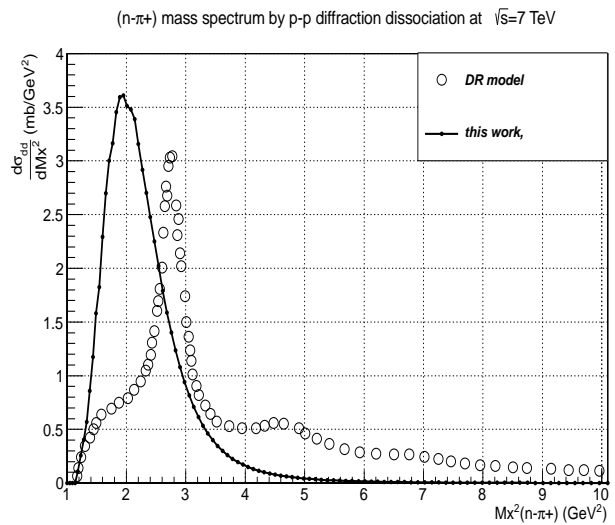}
\caption{ The invariant mass spectrum of the proton-pion system produced in the proton-proton interactions 
at $\sqrt{s}=7$~TeV. The solid line corresponds to calculations according to this work, the points 
are calculations~\cite{jen2012}.}
\label{m7tev}
\end{figure}

\begin{figure}
\includegraphics[height=2.8in,width=3.5in]{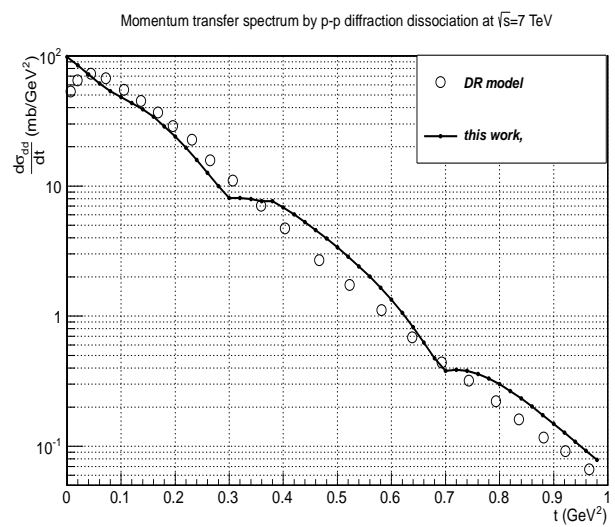}
\caption{ The momentum transfer spectrum of the proton-pion system produced in the proton-proton interactions 
at $\sqrt{s}=7$~TeV. The solid line corresponds to calculations according to this work, the points are 
calculations~\cite{jen2012}.}
\label{t7tev}
\end{figure}


\section{ Diffraction dissociation model}

Fig.~\ref{diagram} shows the quark-diquark (qQ) diagram for the low mass 
(near the mass threshold $\sim$1~GeV for nucleons) diffraction dissociation of the  
projectile nucleon $N$ in the nucleon-nucleon interaction. $M_x^2$ is the invariant mass of the 
dissociated nucleon squared, $t$ is the momentum transfer between nucleons, $\sqrt{s}$ is the total 
energy in the center of mass system of interacting nucleons. The diffraction dissociation 
amplitude $A_{dd}$ corresponding to the diagram reads:
\[
A_{dd}\sim V_{ff}(t_1)G(t_1)A_{el}^{N\pi}(s_1,t),
\]
where $V_{ff}(t_1)=\exp(-\alpha^{\prime}t_1)$ is the vertex form-factor, $G(t_1)$ is the $q\bar{q}$ propagator, and 
$A_{el}^{N\pi}(s_1,t)$ is the elastic amplitude according to the $qQ$-model with one or two 
springy Pomeron $\mathcal{P}$ exchange~\cite{vmg2014}.    
$t_1$ is the momentum transfer squared of the dissociated nucleon and $s_1$ is the total momentum 
squared of elastic scattering. The propagator $G(t_1)$ can be written in the Reggeized 
form~\cite{d1964,t1975}:
\[
G(t_1) = \frac{\exp\left\{\alpha^{\prime}(t_1-\mu^2+i\Gamma)
\displaystyle\ln\left[\frac{M_x^2-M_N^2}{\xi_o}
\right]\right\}}
{\alpha^{\prime}(t_1-\mu^2+i\Gamma)},
\]
where $\mu$ is the pion mass (the efficient mass of quark-antiquark intermediate state), 
$M_N$ is the mass of dissociated nucleon, $\alpha^{\prime}= \ $0.9~GeV$^2$, $\xi_o= \ $1~GeV$^2$ and we 
introduce the imaginary part of the propagator pole, $\Gamma= \ $0.09~GeV$^2$. The 
latter value was chosen from the comparisons of the $M_x$-spectra with measurements.
 
The low mass single diffraction dissociation spectrum reads~\cite{gerhold1976}:
\[
\frac{d\sigma}{dM_x^2}=A\int dt \ d\cos\theta \ d\phi \frac{q_N}{M_x}|A_{dd}|^2,
\]
where $\theta$, $\phi$ and $q_N$ are the polar and azimuthal angles and the magnitude of the 
nucleon momentum in the rest system of $N\pi$ (the Gottfried-Jackson system). $A$ is the phase-space 
normalization constant. The single diffraction dissociation  model was compared with experimental 
data in terms of $d\sigma/dM_x$, 
\[
\frac{d\sigma}{dM_x}=2M_x\frac{d\sigma}{dM_x^2},
\]
the invariant differential cross-sections. The data were taken from~\cite{kerret1976} 
for $p+p\rightarrow (n\pi^{+})+p$ reaction at CERN ISR $\sqrt{s}$= 45 GeV. 
The detector acceptance covers $0<\phi<2\pi$ and different ranges of $\cos\theta$ 
and $|t|$ depending on the analysis of events.

Fig.~\ref{mass1} shows the invariant mass spectrum of the proton-pion system produced 
in the proton-proton interactions at $\sqrt{s}= \ $45~GeV. The histogram corresponds to 
calculations according to this work, the points are experimental data~\cite{kerret1976}. 
Both the data and simulation correspond to -0.3$ \ <\cos\theta< \ $0.3 
and 0.05$ \ <|t|< \ $0.2~GeV$^2$. Fig~\ref{mass2} shows  the 
invariant mass spectrum of the proton-pion system produced in the proton-proton interactions 
at $\sqrt{s}=45$~GeV. The solid histogram corresponds to calculations according to this work, 
the dashed line is the Deck-model calculations from~\cite{gerhold1976}, the points are 
experimental data~\cite{kerret1976}.  Both the data and simulation correspond to 
-0.3$ \ <\cos\theta< \ $1 and 0.05$ \ <|t|< \ $1.2~GeV$^2$. One can see that the proposed model 
shows good agreement with data for the invariant mass spectrum. 

The model predictions for the proton-proton diffraction dissociation at the LHC energies are shown in 
Figs.~\ref{m7tev}-\ref{t7tev}. Fig.~\ref{m7tev} shows  the invariant mass spectrum of 
the proton-pion system produced in the proton-proton interactions at $\sqrt{s}=7$~TeV. 
The solid line corresponds to calculations according to this work, the points are 
calculations according to the dual Reggeon (DR) model~\cite{jen2012}.  Both calculations correspond to 
-1$ \ <\cos\theta< \ $1 and 0$ \ <|t|< \ $1~GeV$^2$. Fig.~\ref{t7tev} shows  the momentum transfer 
spectrum of the proton-pion system produced in the proton-proton interactions 
at $\sqrt{s}=7$~TeV. The solid line corresponds to calculations according to this 
work, the points are calculations~\cite{jen2012}. Both calculations correspond to 
-1$ \ <\cos\theta< \ $1 and 0$ \ <|t|< \ $6~GeV$^2$. It is seen that the mass spectra differ, 
while the momentum transfer spectra are more consistent. Experimental data are needed to 
resolve the observed disagreement.


\begin{figure}
\includegraphics[height=2.8in,width=3.5in]{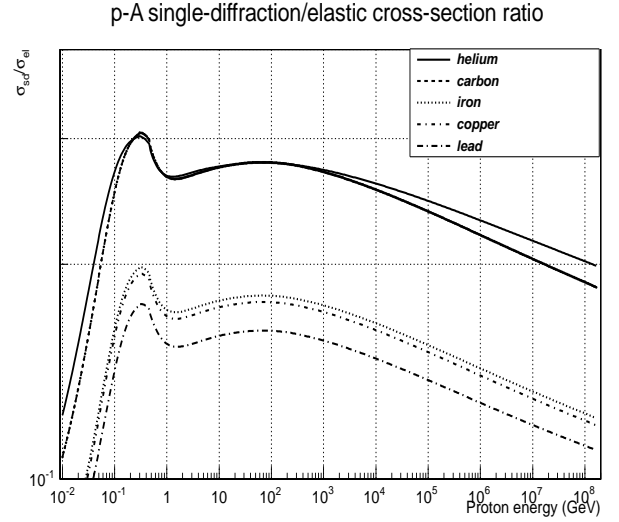}
\caption{The ratio of the single diffraction to elastic cross-section for the proton-nucleus 
interactions versus the proton energy. }
\label{psd}
\end{figure}

\begin{figure}
\includegraphics[height=2.8in,width=3.5in]{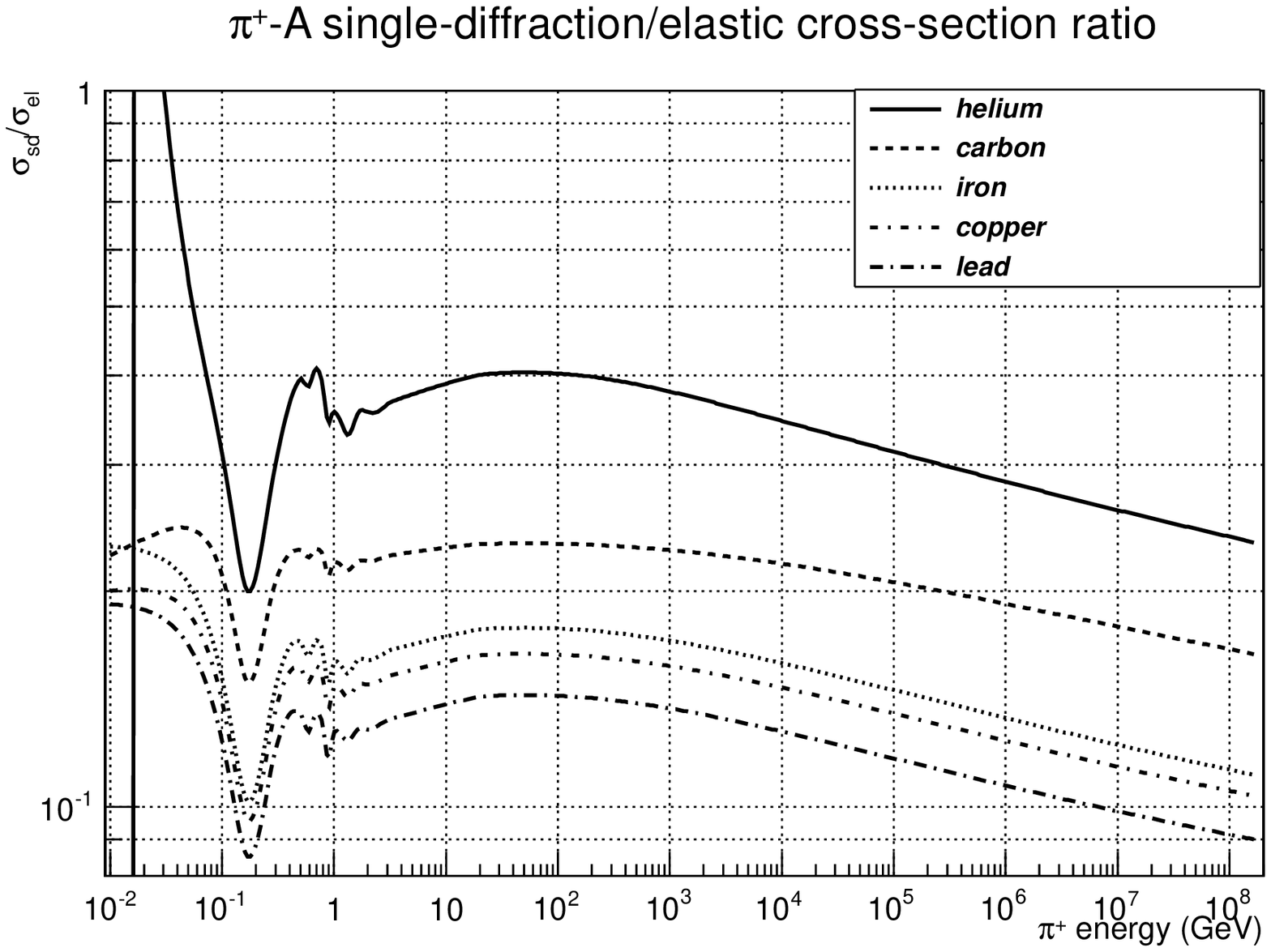}
\caption{The ratio of the single diffraction to elastic cross-section for the pion-nucleus 
interactions versus the pion energy. }
\label{pipsd}
\end{figure}

\begin{figure}
\includegraphics[height=2.8in,width=3.5in]{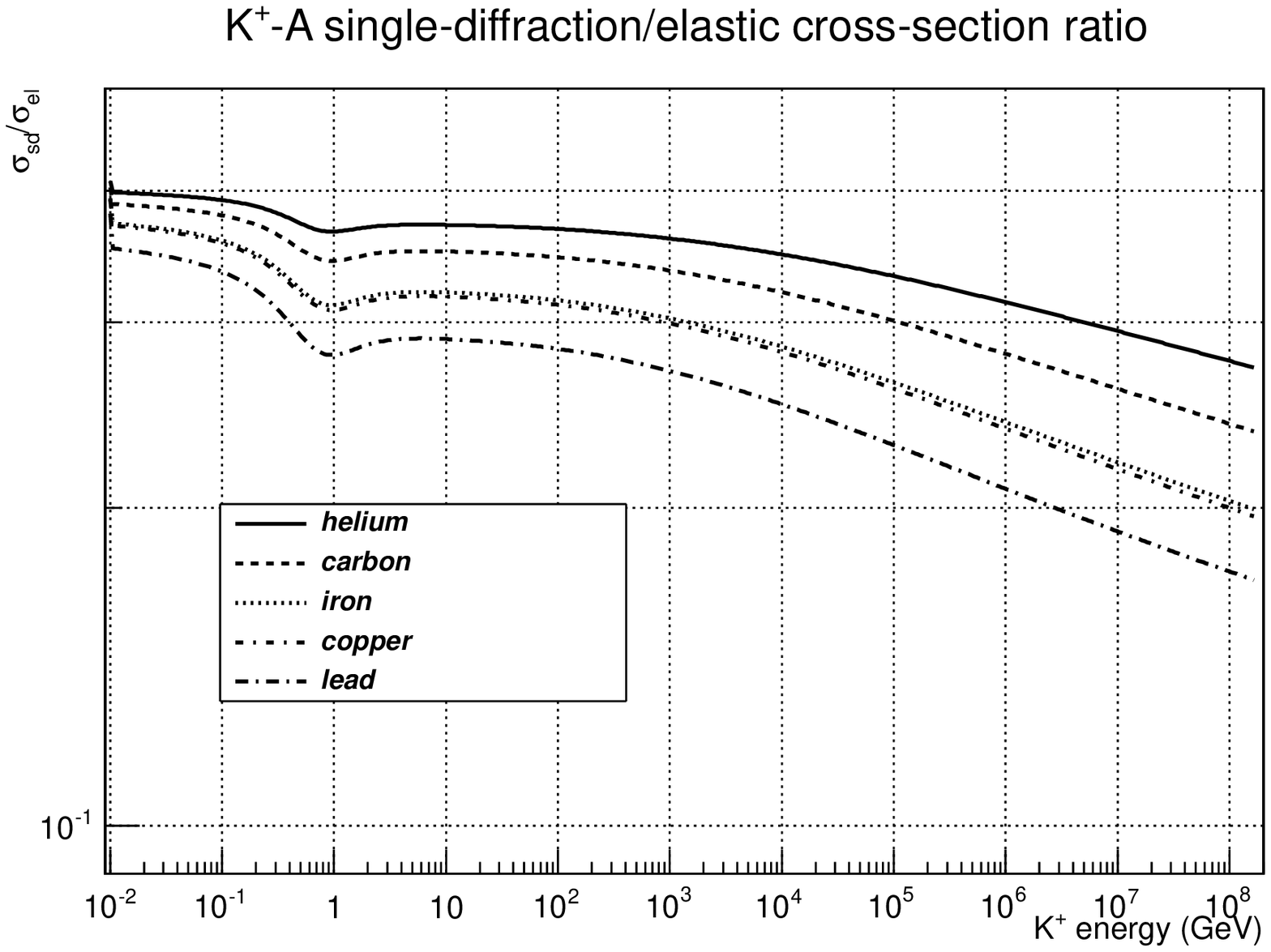}
\caption{ The ratio of the single diffraction to elastic cross-section for the kaon-nucleus 
interactions versus the kaon energy.}
\label{kpsd}
\end{figure}


\begin{figure}
\includegraphics[height=2.8in,width=3.5in]{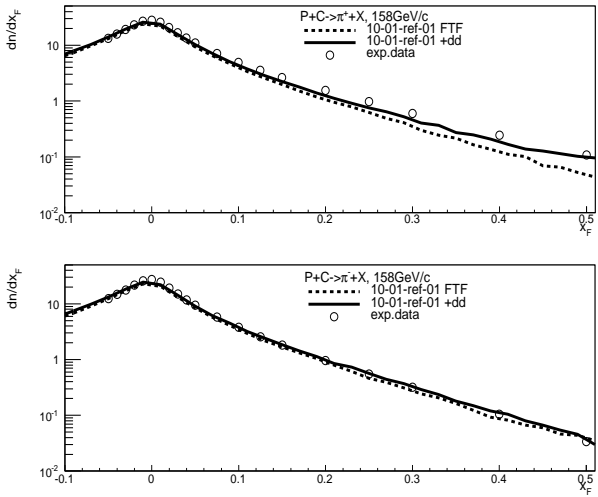}
\caption{The $x_F$-distribution of secondary $\pi^{+}$ (upper) and $\pi^{-}$ (lower) 
produced in the proton-carbon interactions with the proton momentum 
158~GeV/c. The solid line - the FTF model with low mass DD, the dashed line - the FTF model, 
the points are the NA49 data~\cite{na49}. }
\label{xf158}
\end{figure}

\begin{figure}
\includegraphics[height=2.8in,width=3.5in]{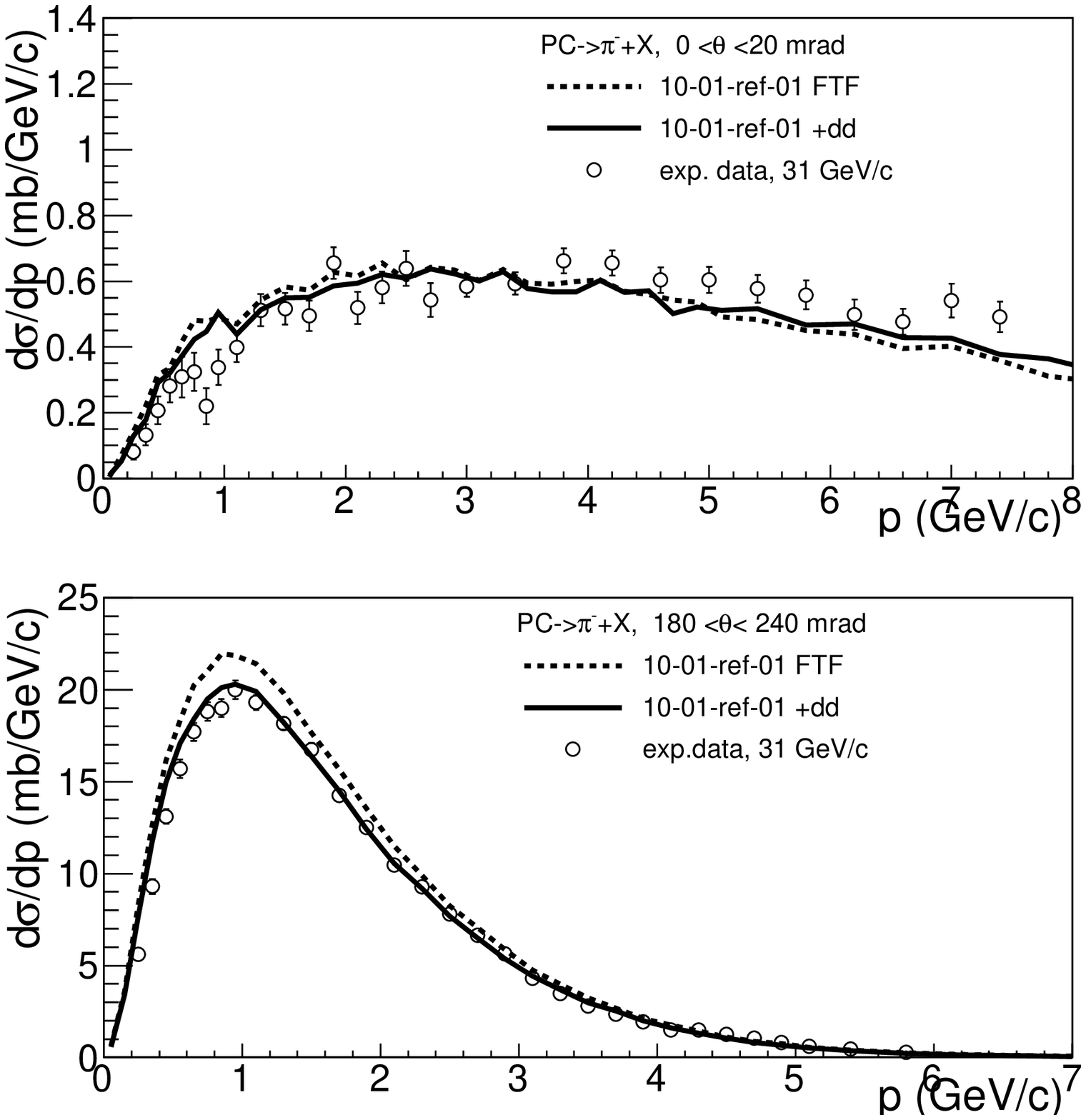}
\caption{The momentum distribution of secondary $\pi^{-}$ produced 
in the proton-carbon interactions with the proton momentum 31~GeV/c. 
The upper plot corresponds to the scattering angles  
0-20 mrad, the lower - 180-240 mrad. 
The solid line - the FTF model with the low mass DD, 
the dashed line - the FTF model, the points are the NA61 data~\cite{na61}. }
\label{pim31}
\end{figure}

\begin{figure}
\includegraphics[height=2.8in,width=3.5in]{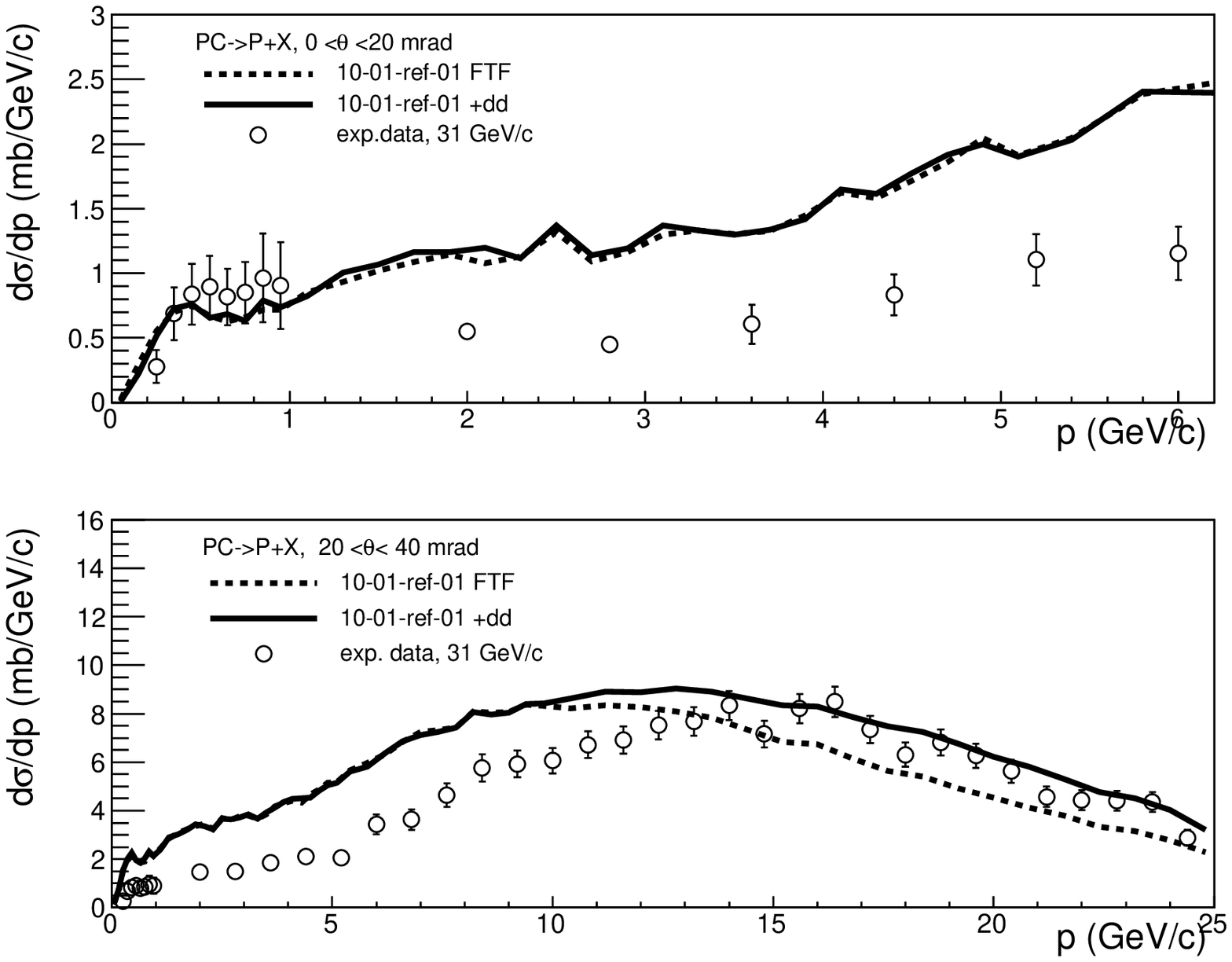}
\caption{The momentum distribution of secondary protons  produced 
in the proton-carbon interactions with the proton momentum 31~GeV/c. 
The upper plot corresponds to the scattering angles in the range 0-20 mrad, 
the lower - 20-40 mrad. The solid line - the FTF model with the low mass DD, 
the dashed line - the FTF model, the points are the NA61 data~\cite{na61}. }
\label{p31}
\end{figure}

\begin{figure}
\includegraphics[height=2.8in,width=3.5in]{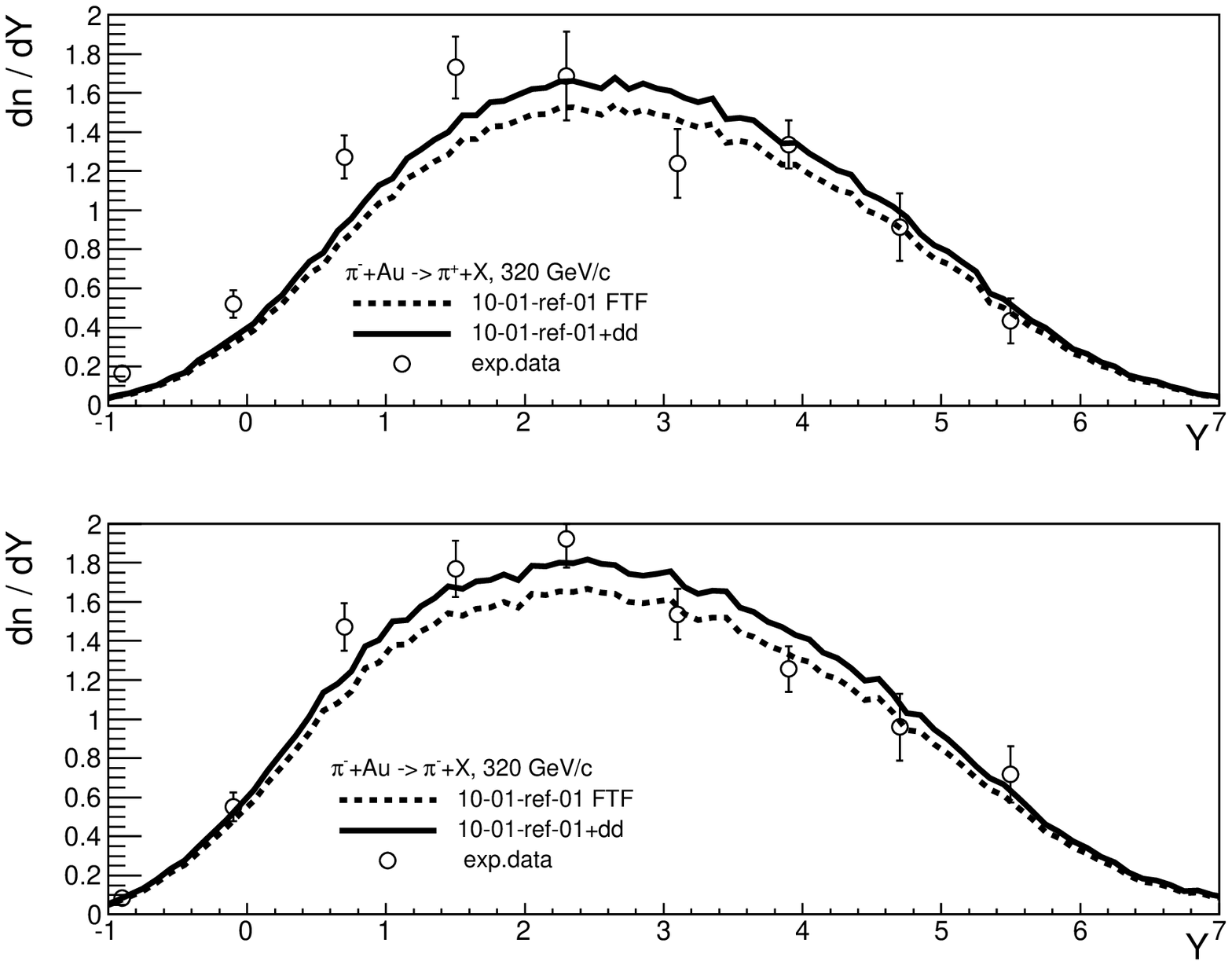}
\caption{The rapidity distribution of secondary $\pi^{+}$ (upper) and $\pi^{-}$ (lower) 
produced in the $\pi^{-}$-gold interactions with the  $\pi^{-}$ momentum 
320~GeV/c. The solid line - the FTF model with the low mass DD, the dashed 
line - the FTF model, the points are the E597 data~\cite{e597}. }
\label{y320}
\end{figure}


\section{ Simulation framework}

The \Gfour simulation tool-kit~\cite{g4} was chosen as the framework for the simulation of 
the low mass diffraction dissociation according to the proposed model. The simulation 
implementation consists of the single diffraction hadron-nucleus cross-section model, 
the model final state generator and the \Gfour application allowing one to 
reproduce approximately the experimental set-ups.

\subsection{ Single diffraction cross-section}

In the framework of simplified Glauber-Gribov model~\cite{vmg2009} the cross 
sections read:
\[
\sigma^{hA}_{tot}=2\pi R^2\ln\left[1+\frac{A\sigma^{hN}_{tot}}{2\pi R^2}\right],
\
\sigma^{hA}_{in} = \pi R^2\ln\left[1+\frac{A\sigma^{hN}_{tot}}{\pi R^2}\right],
\]
\[
\sigma^{hA}_{prod} =
\pi R^2\ln\left[1+\frac{A\sigma^{hN}_{in}}{\pi R^2}\right],\
\sigma^{hA}_{qe}=\sigma^{hA}_{in}-\sigma^{hA}_{prod},
\]
\[
\sigma^{hA}_{sd}(hA\rightarrow XA)=\pi R^2\left\{\alpha-
\ln\left[1+\alpha\right]\right\},
\]
\[
\alpha = \frac{ A\sigma^{hN}_{tot} }{ 2\pi R^2+A\sigma^{hN}_{tot} }.
\]
Where $\sigma^{hA}_{tot}$, $\sigma^{hA}_{in}$, $\sigma^{hA}_{prod}$, 
$\sigma^{hA}_{qe}$ and $\sigma^{hA}_{sd}(hA\rightarrow XA)$ 
are the total, inelastic,  production, quasi-elastic and single-diffraction 
(responsible for low mass DD) cross sections of a hadron on a nucleus A, 
respectively. The total and inelastic hadron-nucleon cross-sections 
are $\sigma^{hN}_{tot}$ and $\sigma^{hN}_{in}$, respectively. R is the nucleus radius. 
In this framework the cross-sections are subdivided as follow:
\[
\sigma^{hA}_{el}=\sigma^{hA}_{tot}-\sigma^{hA}_{in}=\sigma^{hA}_{coh-el}+\sigma^{hA}_{sd}(hA\rightarrow XA),
\]
where $\sigma^{hA}_{coh-el}$ is the coherent elastic cross-section. The Glauber-Gribov cross-sections are 
the part of the \Gfour library, therefore one can introduce the low mass DD generator 
via the ratio $\sigma^{hA}_{sd}(hA\rightarrow XA)/\sigma^{hA}_{el}$. 

\subsection{ Final state generator}

The corresponding generator was implemented in the framework of the \Gfour tool-kit. 
This library allows one to implement the hadron elastic process modified so that it activates the 
low mass DD generator with the probability equal to the $\sigma_{sd}/\sigma_{el}$ ratio. The 
generator utilizes the precalculated model mass spectrum to excite and retard the projectile 
hadron. The  sampled $M_x$ value affects the longitudinal momentum transfer between 
the projectile and the target nucleon. The excited hadron experiences elastic 
scattering on the target nucleon according to the precalculated transverse momentum spectrum. 
A proper excited hadron   
resonance (f.e. for nucleons, these are $N(1440)$, $N(1520)$ and $N(1680)$) 
is selected according to the mass spectrum and the reaction kinematics. Then the 
resonance is decayed recursively down to not short-lived particles. The latter produce the 
final state of the low mass DD generator and are subject of the further \Gfour tracking.

\subsection{ Simulation application}

A \Gfour application was implemented creating simplified geometries of few experiments. 
The geometry includes the volume target with both material and sizes modified from the command line. 
The target was surrounded by the sensitive detector volume were the secondary tracks were 
analyzed in terms of their kinematics. The application physics list allows a user to change 
the active physics, in particular to switch on/off the DD generator.

\section{Comparison with experimental data}

The spectra of secondary particles produced in the inelastic hadron-nucleus interactions 
measured in different thin target experiments we used for the low mass DD generator 
validation.  The \Gfour FTFP\_BERT physics list was used in two modes - with and 
without the low mass DD generator activation.

Fig.~\ref{xf158} shows the $x_F$-distribution of secondary $\pi^{+}$ (upper) and $\pi^{-}$ (lower) 
produced in the proton-carbon interactions with the proton momentum 158~GeV/c. 
The solid line - the FTF model with the low mass DD, the dashed line - the FTF model, 
the points are the NA49 data~\cite{na49}. The longitudinal scaling variable  $x_F$  is defined 
in the center of mass system of the projectile proton and the target nucleon: 
\[
x_F = \frac{2p_L}{\sqrt{s}},
\]
where $p_L$ is the longitudinal secondary particle momentum and $\sqrt{s}$ is the total 
energy of the projectile proton and the target nucleon. It is seen that the low mass DD 
generator increases the spectrum in the region of high  $x_F$, which corresponds to 
forward kinematics region of secondary pion. This correction improves the agreement 
between the measurement and the simulation. 

Fig.~\ref{pim31} shows the momentum distributions of secondary $\pi^{-}$ produced 
in the proton-carbon interactions with the proton momentum 31~GeV/c. 
The upper plot corresponds to the scattering angles  0-20 mrad, the lower - 180-240 mrad. 
The solid line - the FTF model with low mass DD, the dashed line - the FTF model, 
the points are NA61 data~\cite{na61}. One can see that the low mass DD correction improves 
the agreement with the experiment especially in the scattering angle range 180-240 mrad. 
Fig.~\ref{p31} shows the momentum distributions of secondary protons  produced 
in the proton-carbon interactions with the proton momentum 31~GeV/c. 
The upper plot corresponds to the scattering angles in the range 0-20 mrad, 
the lower - 20-40 mrad. The solid line - the FTF model with the low mass DD, 
the dashed line - the FTF model, the points are the NA61 data~\cite{na61}. The low mass 
DD correction improves essentially the agreement with the data in the angle range 20-40 mrad. 

Fig~\ref{y320} shows the rapidity distributions of secondary $\pi^{+}$ (upper) and $\pi^{-}$ (lower) 
produced in the $\pi^{-}$-gold interactions with the $\pi^{-}$ momentum 320~GeV/c. 
The solid line - the FTF model with the low mass DD, the dashed line - the FTF model, 
the points are the E597 experiment data~\cite{e597}. The rapidity of secondary pions corresponds to the 
laboratory frame. The low mass DD activation results in better agreement with the data.

\section{Summary}

The low mass single diffraction dissociation model based on the quark-diquark 
representation of hadron-nucleon interaction is proposed. The mass spectra of the nucleon-pion system are 
compared with experimental data and the predictions of other models. It was found better agreement 
with the data compared to the DHD model. The reason is that the proposed model utilizes the modified 
propagator $G(t_1)$ with increased imaginary part due to introduction of the $\Gamma$-parameter. 
The  hadron-nucleus single-diffraction cross-sections are calculated in the framework of the 
Glauber-Gribov model for integral cross-sections. The model predictions are compared with the experimental 
data for the  different distributions of secondary particles produced in hadron-nucleus 
interactions in the momentum range 31-320 GeV/c. The comparisons show that the DD generator improves 
the agreement with the experimental data.  The model and the generator tuning as well as extended 
comparisons with experimental data are in progress and will be reported elsewhere.

\section*{Acknowledgment}
 
The author is thankful to S. Bertolucci, S. Giani and M. Mangano for stimulating 
discussions and support. V. Ivanchenko and A. Ribon presented the interface for 
the model introducing in the framework of \Gfour.



\end{document}